\documentstyle[12pt]{article}

\begin{document}

\baselineskip 22pt

\noindent
\hspace*{11.6cm}
KEK-TH-492\\
\noindent
\hspace*{11.3cm}
SNUTP 96-083\\
\noindent
\hspace*{11.5cm}
YUMS 96-017\\
\noindent
\hspace*{11.4cm}
(March 1997)\\

\begin{center}

{\Large \bf Average Kinetic Energy of Heavy Quark and 
Virial Theorem\\}

\vspace{0.2cm}

Dae Sung Hwang$^1$, ~C. S. Kim$^{2,3}$ ~and~ Wuk Namgung$^4$
\end{center}

\noindent Department of Physics, Sejong University,
Seoul 143-747, Korea\footnote{dshwang@phy.sejong.ac.kr}\newline
\noindent Department of Physics, Yonsei University,
Seoul 120-749, Korea\footnote{kim@cskim.yonsei.ac.kr}\newline
\noindent Theory Division, KEK, Tsukuba,
Ibaraki 305, Japan\footnote{cskim@kekvax.kek.jp}\newline
\noindent Department of Physics, Dongguk University,
Seoul 100-715, Korea\footnote{ngw@cakra.dongguk.ac.kr}

\begin{center}
(\today)
\vspace{0.5cm}

{\bf Abstract}
\end{center}

We derive the virial theorem of the relativistic two-body system
for the study of the $B$-meson physics.
It is also shown that the solution of the variational equation always 
satisfies the virial theorem. From the virial theorem
we also obtained
$\mu_\pi^2 \equiv -\lambda_1 \equiv \langle {\bf p}^2\rangle  =
0.40\sim 0.58$ GeV$^2$,
which is consistent with the result 
of the QCD sum rule calculations of Ball $et$ $al.$

\vspace{1.0cm}

\vfill

\noindent PACS codes: 12.39.Hg, 12.39.Ki, 12.39.Pn, 14.40.Nd\\
Key words: average kinetic energy, virial theorem, $B$-meson,
relativistic quark model, heavy quark effective theory,
variational method, potential model

\thispagestyle{empty}
\pagebreak
\baselineskip 22pt

\noindent
{\bf \large I. Introduction}\\

The $B$-physics provides many important and attractive physics informations,
and is under active experimental and theoretical investigations.
The $B$-factories at KEK and SLAC will give valuable clues for the better
understanding of the Standard Model (SM) and the beyond.
We expect the $B$-meson system will show the CP-violation 
phenomena \cite{sanda}, for which we have only the 
$K_L\rightarrow \pi\pi$ decay for more than 30 years.
The mechanism of CP-violation through the complex phase of the
Kobayashi-Maskawa three family mixing matrix
\cite{ckm} in the Weinberg-Salam
model is presently the CP-violation within the SM.
In order to understand and precisely test the SM, it is essential
to know the values of the Kobayashi-Maskawa matrix elements,
especially through the determinations  of $V_{ub}$ \cite{vub}
and $V_{td}$, 
and confirm the unitarity triangle \cite{quinn},
both of which will be best probed at the forthcoming $B$-factories.

Recently, it has been an important subject to obtain an accurate value
of the kinetic energy, 
$\mu_\pi^2~(\equiv -\lambda_1 \equiv \langle {\bf p}^2 \rangle)$,
of the heavy quark inside $B$-meson in connection with the heavy quark 
effective theory (HQET) \cite{isgur,neubert}.
Ball $et$ $al.$ \cite{ball} calculated $\mu_\pi^2$ using the QCD sum rule
approach and obtained $\mu_\pi^2= 0.50 \pm 0.10$ ${\rm GeV}^2$
for $B$-meson, while Neubert \cite{nbt3} obtained 
$-\lambda_1= 0.10 \pm 0.05$ ${\rm GeV}^2$.
Neubert also derived \cite{neubert2}
the field-theory version of the virial theorem within the HQET framework.
Based on the theorem, he implied that the result 
$\mu_\pi^2 \sim 0.5\ {\rm GeV}^2$
of the QCD sum rule calculations of Ball $et$ $al.$ is too large.
However, it should be noted that Refs. \cite{ball} and \cite{nbt3}
differ in the choice of the 3-point correlation functions used to
estimate the matrix elements of interest.
The difference in the numerical values obtained in these two
calculations is understood in terms of the contributions of
excited states, which in principle must be subtracted in any
QCD sum rule analysis.
In practice, this subtraction can only be done approximately.
Therefore, the numerical differences between Refs. \cite{ball} and
\cite{nbt3} indicate the limited accuracy of the QCD sum rule
approach.
Bigi $et$ $al.$ \cite{bigi123} derived an inequality between the
expectation value of the kinetic energy of the heavy quark
inside the hadron and that of the chromomagnetic operator,
$\langle {\bf p}^2 \rangle \,\ge\, {3 \over 4} ({M_V}^2-{M_P}^2)$,
which gives $\mu_\pi^2 \ge 0.36\ {\rm GeV}^2$ for $B$-meson system.
However,
Kapustin $et$ $al.$ \cite{kapustin} showed later that this lower bound
could be significantly weakened by higher order perturbative corrections.
Previously we also calculated \cite{hkn} the value of
$\langle {\bf p}^2 \rangle$ by applying the variational method 
to the relativistic Hamiltonian, 
and obtained $\langle {\bf p}^2 \rangle = 0.44$ GeV$^2$. 
Similarly de Fazio \cite{Fazio} computed the matrix elements of the
kinetic energy operator by means of a QCD relativistic potential
model, and found $\mu_\pi^2 = 0.46$ GeV$^2$.

Besides the above theoretical calculations of $\mu_\pi^2$,
Gremm $et$ $al.$ \cite{gremm} extracted the average kinetic energy
by comparing the prediction of the HQET \cite{bigi} with
the shape of the inclusive $B \rightarrow X l {\nu}$
lepton energy spectrum \cite{cleo93b} for $E_l \ge 1.5$ GeV,
in order to avoid the contamination from the secondary leptons of cascade
decays of $b \rightarrow c \rightarrow s l \nu$.
They obtained
$-\lambda_1 = 0.19 \pm 0.10\ {\rm GeV}^2$.
Combining the experimental data on the inclusive decays of 
$D \rightarrow X e \nu$, $B \rightarrow X e \nu$ and
$B \rightarrow X \tau \nu$,
Ligeti $et$ $al.$ \cite{Ligetiexp} derived the bound of 
$\mu_\pi^2 \le 0.63$ GeV$^2$ if $\bar \Lambda \ge 0.240$ GeV, or
$\mu_\pi^2 \le 0.10$ GeV$^2$ if $\bar \Lambda \ge 0.500$ MeV.
Li $et$ $al.$ \cite{Li} obtained the value of $-\lambda_1$
centered at $0.71$ GeV$^2$ from the
analysis of the inclusive radiative decay
$B \rightarrow X_s \gamma$ \cite{Liexp} within the perturbative
QCD framework.
Related with the comparison of various theoretical calculations of
$\mu_\pi^2$, we note that Ref. \cite{neu4} emphasizes that one has to
be careful when comparing the values of $-\lambda_1$ obtained using
different theoretical methods.
For instance, QCD sum rule determinations of $-\lambda_1$ for the
ground state heavy mesons and baryons are affected by a renormalon ambiguity
problem \cite{neu4}.

Concerned with the phenomenological importance of the numerical value for 
the kinetic energy of the heavy quark,
we would like to derive the virial theorem of the  two-body system within 
the relativistic potential model approach
for the study of the $B$-meson system.
We show at the end  that $\mu_\pi^2 \sim 0.50\ {\rm GeV}^2$ of Ball $et~al.$ 
\cite{ball} is  consistent with our virial theorem result.
In Section II we present the variational analysis of the relativistic
quark model for $B$-meson system, and compare it with the HQET.
In Section III we derive the virial theorem of the relativistic 
two-body system which is appropriate for $B$-meson system,
and we show that the solutions of the variational 
equation always satisfy the virial theorem automatically.
Section  IV contains discussions and  conclusions.
\\

\noindent
{\bf \large II. Relativistic Quark Model}\newline

For the study of the bound state properties of hadrons which contain
both heavy and light quarks like $B$-meson, it is appropriate to
use the relativistic quark model \cite{hkn,lucha1},
which is a potential model approach with relativistic kinematics.
In the relativistic quark model, the Hamiltonian for $B$-meson is
given by
\begin{equation}
H={\sqrt{{\bf p}^2+M^2}}+{\sqrt{{\bf p}^2+m^2}}+V(r)
\approx
M+{{{\bf p}^2}\over {2M}}+{\sqrt{{\bf p}^2+m^2}}+V(r)
\label{a1}
\end{equation}
in the $B$-meson rest frame,
where $M$ and $m$ are heavy and light quark mass respectively.
For the potential in (\ref{a1}), we use the Cornell potential
\cite{lucha1,eich,quigg},
\begin{equation}
V(r)=-{{{\alpha}_c}\over {r}}+Kr+V_0,\qquad
\alpha_c\equiv {4\over 3} \alpha_s ~,\qquad V_0= {\rm constant} ~.
\label{a2}
\end{equation}
It is difficult to solve the eigenvalue equation of the Hamiltonian
operator (\ref{a1}), so we use the variational method.
The variational method is particularly useful in the present work,
since we will show in the following section 
that the solutions of the variational equation
always satisfy the virial theorem. % and we will use this property.

In the variational method,
the expectation value of the Hamiltonian is calculated
with some trial wave function
which has a variational parameter.
The value of the variational parameter
is determined by the stationary condition
(variational equation or so-called gap equation).
We take the variational parameter which has the dimension of mass.
Then from the dimensional 
analysis, the expectation value of each term in the Hamiltonian 
can be expressed as
\begin{eqnarray}
\langle {\bf p}^2 \rangle  &=&
C \mu^2,
\nonumber\\
\langle \sqrt{{\bf p}\,^2+ m^2} \rangle  &=&
a_1\mu +a_2\, m^2/\mu +O\Bigl( (m^2/\mu)^2\Bigr) ,
\nonumber\\
\langle -{{{\alpha}_c}\over {r}}+Kr+V_0 \rangle  &=&
\alpha_c (-b_1\mu )+ K (b_2 / \mu ) + V_0,
\label{a3}
\end{eqnarray}
where $\mu$ is the variational parameter of mass dimension,
and $C$, $a_1$, $a_2$, $b_1$ and $b_2$ are
dimensionless numerical constants.
Collecting the terms in (\ref{a3}), 
we have
\begin{eqnarray}
\langle H \rangle_{\mu} =E(\mu )&=&
M+[V_0+(a_1-b_1\alpha_c)\mu +(a_2m^2+b_2K)/\mu ]+{C\over 2M}\mu^2
\nonumber\\
&= &
M+[V_0+\beta\mu +\gamma /\mu ]+{C\over 2M}\mu^2 ,
\label{a4}
\end{eqnarray}
where
\begin{equation}
\beta\equiv a_1-b_1\alpha_c\qquad {\rm and}\qquad
\gamma\equiv a_2m^2+b_2K.
\label{a4a}
\end{equation}
We neglected $O\Bigl( (m^2/\mu)^2\Bigr)$ terms in (\ref{a4}).
Then the variational equation reads
\begin{equation}
{\partial\over \partial\mu}E(\mu )
=\beta -\gamma /\mu^2 +{C\over M}\mu =0.
\label{a5}
\end{equation}

Rather than solving this equation numerically,
we obtain the solution as a power
series in $1/M$, since $M$ is large.
Noting that the solution is given by
$\bar{\mu} \sim {\sqrt{\gamma /\beta }}$
for very large value of $M$, we expand
\begin{equation}
\bar{\mu} =h_0+h_1{1\over M}+h_2{1\over M^2}+\cdots .
\label{a6}
\end{equation}
Substituting (\ref{a6}) into (\ref{a5}) and matching order by order, we get
\begin{equation}
h_0=\sqrt{{\gamma\over\beta }},\,\,\,\,\,
h_1=-{C\over 2}\Bigl( {\gamma\over {\beta}^2}\Bigr),\,\,\,\,\,
h_2={5C^2\over 8}\sqrt{{\gamma\over\beta }}
\Bigl( {\gamma\over {\beta}^3}\Bigr),\,\,\,\,\,\cdots.
\label{a7}
\end{equation}
Then we get $E(\bar{\mu} )$, the $B$-meson mass $M_B$,
as a power series in $1/M$,
\begin{equation}
M_B=E({\bar{\mu}} )=
M+\Bigl( V_0+2\sqrt{\gamma\beta }\,\Bigr) +
{C\over 2}\Bigl( {\gamma\over\beta }\Bigr) {1\over M}
+O({1\over M^2}).
\label{a8}
\end{equation}
Therefore we have performed the $1/M$ expansion for $M_B$ in the
relativistic quark model in the framework of the variational method.
Then, let us compare
the series expansion in (\ref{a8}) with that of the HQET
\cite{isgur,neubert} which is written as
\begin{equation}
M_B=M+{\bar{\Lambda}}+{1\over 2M}(T+{\nu}_B\Omega )
+O({1\over M^2}),
\label{a9}
\end{equation}
where ${\nu}_B=1/4$ and $-3/4$ for vector and pseudoscalar $B$-meson
respectively.
The zeroth order term ${\bar{\Lambda}}$
in (\ref{a9}) which is the contribution from the 
light degrees of freedom corresponds to the sum of the
expectation value of the light quark kinetic energy and that of the 
potential energy, 
\begin{equation}
{\bar{\Lambda}}\longleftrightarrow 
\langle {\sqrt{{\bf p}^2+m^2}}+V(r)\rangle =
V_0+2{\sqrt{\gamma\beta}}.
\label{a10}
\end{equation}
The heavy quark kinetic energy term $T$ in (\ref{a9})
has the correspondence
\begin{equation}
T\longleftrightarrow 
\langle {\bf p}^2\rangle =C\, {\gamma\over\beta}\, .
\label{a11}
\end{equation}
If we had included a spin dependent potential which is inversely 
proportional to the heavy quark mass, the expectation value of which 
would have corresponded to the chromomagnetic interaction term in (\ref{a9})
as
\begin{equation}
{\nu_B\Omega\over 2M} \longleftrightarrow 
\langle V_s\rangle ={1\over M}\langle v_s\rangle \qquad
{\rm with}\qquad
v_s={2\over {3m}}{\bf s}_1\cdot {\bf s}_2
{\nabla}^2(-{\alpha_c\over r}),
\label{a12}
\end{equation}
where $v_s$ is the nonrelativistic spin-spin interaction potential.
\\

\noindent
{\bf \large III. Virial theorem}\\

In classical mechanics, the virial theorem for a system with
kinetic energy $K({\bf p})$ and
potential energy $V({\bf r})$ is based on the relation
\begin{equation}
{d\over dt}({\bf r}\cdot {\bf p})=
{\bf v}\cdot {\bf p} + {\bf r}\cdot {d\over dt}{\bf p} =
({\partial \over \partial {\bf p}}K({\bf p}))\cdot {\bf p}
+ {\bf r}\cdot (-\nabla V({\bf r})),
\label{a16}
\end{equation}
where $K({\bf p})$ is the kinetic energy given by
$K({\bf p})=\int {\bf F}\cdot d{\bf r} =\int {\bf v}\cdot d{\bf p}$.
The time average of the left hand side of (\ref{a16}) vanishes for
a periodic motion or a bounded motion in an infinite time interval,
hence we get the virial theorem,
\begin{equation}
\langle {\bf p}\cdot {\partial \over \partial {\bf p}}K({\bf p})-
{\bf r}\cdot {\partial \over \partial {\bf r}}V({\bf r})
\rangle_{{\rm time\ average}}=0.
\label{a18}
\end{equation}
This theorem holds for both nonrelativistic and relativistic
kinematics which have the following relations respectively,
\begin{eqnarray}
&&{\bf p}=m{\bf v},\qquad K={{\bf p}^2\over 2m},
\qquad {\bf p}\cdot {\partial K\over \partial {\bf p}}={{\bf p}^2\over m}
=2K;
\label{a19}\\
&&{\bf p}={m{\bf v}\over {\sqrt{1-v^2}}},\qquad
K={\sqrt{{\bf p}^2+m^2}},\qquad
{\bf p}\cdot {\partial K\over \partial {\bf p}}=
{{\bf p}^2\over {\sqrt{{\bf p}^2+m^2}}}.
\nonumber
\end{eqnarray}

In quantum mechanics, the virial theorem for a system with 
Hamiltonian $H=K({\bf p})+V({\bf r})$ is based on the relation
\begin{equation}
{d\over dt}\langle {\bf r}\cdot {\bf p}\rangle =
{1\over i{\not{h}} }\langle [{\bf r}\cdot {\bf p}\, ,\, H]\rangle =
{1\over i{\not{h}} }
\langle {\bf r}\cdot (-{\partial V\over \partial {\bf r}})+
({\partial K\over \partial {\bf p}})\cdot {\bf p}\rangle .
\label{a20}
\end{equation}
For stationary states represented by eigenfunctions of $H$,
the left hand side of (\ref{a20}) is zero.
Hence we get the virial theorem written as
\begin{equation}
\langle {\bf r}\cdot (-{\partial V\over \partial {\bf r}})+
({\partial K\over \partial {\bf p}})\cdot {\bf p}\rangle =0.
\label{a20a}
\end{equation}

However, the functions
which satisfy the virial theorem (\ref{a20a})
are not restricted to the eigenfunctions of $H$.
The solutions of the variational equation also satisfy the theorem
(\ref{a20a}).
With the variational parameter $\mu$ of the mass
dimension, the expectation value of the Hamiltonian is 
expressed as
\begin{equation}
E(\mu )=\langle \psi (\mu )|H|\psi (\mu )\rangle 
=\langle K(p)+V(r)\rangle_{\mu},
\label{a21}
\end{equation}
for central potential $V=V(r)$.
In general, the kinetic and potential energy functions in (\ref{a21})
can be expanded in the Laurent series as
\begin{equation}
K(p)=\sum_nk_np^n,\qquad
V(r)=\sum_nv_nr^n.
\label{a22}
\end{equation}
The expectation values of each terms are written in terms of $\mu$,
\begin{equation}
\langle p^n\rangle_\mu =a_n\, \mu^n,\qquad
\langle r^n\rangle_\mu =b_n\, \mu^{-n},
\label{a23}
\end{equation}
where $a_n$, $b_n$ are numerical constants which have no dimensions.
The expressions in (\ref{a23}) satisfy the
relations,
\begin{equation}
\mu{\partial\over\partial\mu}\langle p^n\rangle_{\mu} 
=\langle p{d\over dp}p^n\rangle_{\mu},\qquad
\mu{\partial\over\partial\mu} \langle r^n\rangle_{\mu} 
=-\langle r{d\over dr}r^n\rangle_{\mu}.
\label{a24}
\end{equation}
{}From (\ref{a21})--(\ref{a24}), we get
\begin{equation}
\mu{\partial\over\partial\mu}E(\mu ) =\langle p{d\over dp}K(p)
-r{d\over dr}V(r)\rangle .
\label{a25}
\end{equation}
The equation (\ref{a25}) means that the variational equation,
$\partial E(\mu )/\partial\mu = 0$,
is equivalent to the virial theorem (\ref{a20a}). If we have used 
the parameter $\mu$
having the dimension of length, we would have gotten the same result except 
for the overall sign, which is irrelevant.

We can also show the relations more explicitly.
For the expectation value of the potential,
\begin{equation}
\langle V(r)\rangle_{\mu}=\int d^3r\, \mid \psi ({\bf r};\,\mu ){\mid}^2 V(r)
=\mu^3 \int dr\, r^2 A(\mu r) V(r),
\label{a26}
\end{equation}
where we defined $A(\mu r)$ such that it satisfies
\begin{equation}
\int d^3r\, \mid \psi ({\bf r};\,\mu ){\mid}^2=
\int dr\, r^2 \mid R (r;\,\mu ){\mid}^2=
\mu^3 \int dr\, r^2 A(\mu r)=1.
\label{a27}
\end{equation}
Changing the integration variable in (\ref{a26}) by $x=\mu r$,
\begin{equation}
\langle V(r)\rangle_{\mu}=\int dx\, x^2\, A(x) V(x/\mu ),
\label{a28}
\end{equation}
then we get the relation
\begin{eqnarray}
\mu{\partial\over\partial\mu}\langle V(r)\rangle_{\mu}&=&
\int dx\, x^2 A(x) \{ -{x\over\mu } V' (x/\mu ) \}
\nonumber\\
&=&\mu^3\int dr\, r^2 A(\mu r)\{ -rV'(r)\}
\nonumber\\
&=&-\langle r{d\over dr}V(r)\rangle ,
\label{a29}
\end{eqnarray}
For the expectation value of the kinetic energy, we
can follow a similar procedure in the momentum space to get the desired 
result
$\mu{\partial\over\partial\mu}
\langle K(p)\rangle_{\mu}=\langle p{d\over dp}K(p)\rangle $.
Therefore we obtain the relation (\ref{a25}) again.

In the above, we have shown that $\psi (\bar{\mu} )$ with $\bar{\mu}$
being the solution of the variational equation,
$\partial E(\mu )/\partial\mu = 0$, automatically satisfies the virial
theorem given by
\begin{equation}
\langle\psi (\bar{\mu} )|p{\partial\over\partial p}K(p)|\psi 
(\bar{\mu} )\rangle =
\langle\psi (\bar{\mu} )|r{\partial\over\partial r}V(r)|\psi 
(\bar{\mu} )\rangle ,
\label{a30}
\end{equation}
even though it is not an eigenfunction of $H$. We emphasize that the 
solution set of the virial theorem contains the functions determined 
by the variational method as well as the eigenfunctions of the 
Hamiltonian.
Repko $et$ $al.$ \cite{repko} previously proved that variational principle
implies the virial theorem,
and Lucha $et$ $al.$ \cite{lucha} derived the relativistic form of the
virial theorem.
Let us show explicitly
that the virial theorem (\ref{a30}) is satisfied for the system given by
the Hamiltonian of (\ref{a1}) and (\ref{a2}).
For $B$-meson which has the Hamiltonian (\ref{a1}) with
$K({\bf p})={\bf p}^2/(2M)+{\sqrt{{\bf p}^2+m^2}}$,
the virial theorem (\ref{a30}) is written as
\begin{equation}
\langle{{\bf p}^2\over M}+{{\bf p}^2\over {\sqrt{{\bf p}^2+m^2}}}\rangle =
\langle r{\partial\over\partial r}V(r)\rangle .
\label{a13}
\end{equation}
The left hand side of (\ref{a13}), which is related with
the kinetic energy, can be calculated from (\ref{a3}),
\begin{eqnarray}
\langle{{\bf p}^2\over M}+{{\bf p}^2\over {\sqrt{{\bf p}^2+m^2}}}\rangle &=&
{C\mu^2\over M}+\langle{\sqrt{{\bf p}^2+m^2}}
-{m^2\over {\sqrt{{\bf p}^2+m^2}}}\rangle 
\nonumber\\
&=&{C\mu^2\over M}+(a_1\mu +a_2\, m^2/\mu )
-2m^2{\partial\over\partial (m^2)}(a_1\mu +a_2\, m^2/\mu )
\nonumber\\
&=&{C\mu^2\over M}+(a_1\mu -a_2\, m^2/\mu ).
\label{a14}
\end{eqnarray}
The right hand side of (\ref{a13}), which is related to
the potential energy, is also given from (\ref{a3}),
\begin{equation}
\langle r{\partial\over\partial r}V(r)\rangle =
\langle{{{\alpha}_c}\over {r}}+Kr\rangle ={\alpha}_c(b_1\mu )+K(b_2/\mu ).
\label{a15}
\end{equation}
Putting (\ref{a14}) and (\ref{a15}) in (\ref{a13}) and dividing by $\mu$,
we get the
variational equation (\ref{a5}).
Therefore we have shown that the virial theorem is satisfied for
the solution of the variational equation of the Hamiltonian
given by (\ref{a1}) and (\ref{a2}).
\\

\noindent
{\bf \large IV. Discussions and Conclusions}\\

We have shown that the relativistic quark model combined with the 
variational method can give many useful results.
The solutions of the variational equation always satisfy 
the virial theorem automatically, 
and the $B$-meson mass $M_B$ is expressible as a power series in $1/M$.
Combining (\ref{a13}) and (\ref{a15}), we get the relativistic virial 
theorem within the Cornell potential,
\begin{eqnarray}
\langle{{\bf p}^2\over M}+{{\bf p}^2\over {\sqrt{{\bf p}^2+m^2}}}\rangle &=&
\langle {\bf r}\cdot \nabla V\rangle \nonumber\\
&=&\langle{{{\alpha}_c}\over {r}}+Kr\rangle ={\alpha}_c(b_1\mu )+K(b_2/\mu ).
\label{cs1}
\end{eqnarray}
With the usual input values \cite{lucha1,eich,quigg} 
of ${\alpha}_s=0.21\sim 0.36$, $K=0.19\ {\rm GeV}^2$, $m=0.15$ GeV,
we obtain 
$\mu_\pi^2   =0.40\sim 0.58\ {\rm GeV}^2$ for Gaussian trial functions,
which is consistent with the result $\mu_\pi^2 \sim 0.50\ {\rm GeV}^2$
of the QCD sum rule calculations of Ball $et$ $al.$ \cite{ball}.
If we use exponential trial functions, we get higher values of
$\mu_\pi^2$.
If we rather had used the virial theorem of the
nonrelativistic two-body system
\begin{equation}
\langle {{\bf p}^2\over M}+{{\bf p}^2\over m}\rangle =
\langle {\bf r}\cdot \nabla V\rangle =
\langle {\alpha_c\over r}+Kr\rangle ,
\label{a33}
\end{equation}
we would have obtained the value  
$\mu_\pi^2 = 0.10\sim 0.15$ GeV$^2$, which is very small compared with
the above relativistic result. 

If we apply the virial 
theorem to the system of one-body in an external potential, it reads
\begin{equation}
\langle{{\bf p}^2\over M}\rangle =
\langle r{\partial\over\partial r}V(r)\rangle ,
\label{a31}
\end{equation}
with the nonrelativistic kinematics for the heavy quark.
This corresponds to a one-body (heavy quark) motion in a fixed external 
potential of the background system which produces the potential 
in the meson rest frame. 
Compared with the hydrogen atom, where the nonrelativistic one-body
virial theorem (through the reduced mass)
is also applied, the roles of the heavy (proton) and 
light (electron) degrees of freedom are reversed.
Within the potential model approach, 
the background system corresponds to a valence
light quark and virtual gluons, the former carries the compensating
momentum against the heavy quark motion
and the latter contributes as a potential energy. Considering
these, the correct virial theorem for $B$-meson system
within the potential model approach should be the form of the
two-body closed system as given in (\ref{cs1}).

In conclusion, we derived the virial theorem for the relativistic
two-body system, and numerically obtained
$\mu_\pi^2 \equiv -\lambda_1 \equiv \langle {\bf p}^2\rangle  =
0.40\sim 0.58$ GeV$^2$,
which is consistent with the result 
of the QCD sum rule calculations of Ball $et$ $al.$
It is also shown that the solution of the variational equation always 
satisfies the virial theorem. 
As our final comment, we note the recent observation \cite{neu4}
on the mixing of the operator for the heavy-quark kinetic energy
with the identity operator, which implies that the parameter 
$\lambda_1$ of the heavy-quark effective theory
is not directly a physical quantity, but requires a non-perturbative
subtraction. 
Concerned with the phenomenological importance of the numerical value for 
the kinetic energy of the heavy quark, it would be the most urgent to
try all the possible attempts on the physical understanding of the
kinetic energy of the heavy quark inside $B$-meson system.
\\

%\vfill\eject

\noindent
{\bf  Acknowledgements}\\
 
\indent
The work  was supported
in part by the KOSEF,
Project No. 951-0207-008-2,
in part by the CTP of SNU,
%in part by Yonsei University Faculty Research Grant,
in part by Non-Directed-Research-Fund, Korea Research Foundation 1996,
%in part by Daeyang Foundation at Sejong University,
in part by the BSRI Program, Project No. BSRI-97-2425, and
in part by COE fellowship of the Japanese Ministry of Education, Science
and Culture.

\pagebreak

\end{document}